\documentclass[aps,prl,reprint,showpacs,footinbib,amsmath,amssymb,amsfonts,floatfix,superscriptaddress]{revtex4-1}

\usepackage{graphicx}
\usepackage{epsfig}
\usepackage{color}

\begin{document}

%\preprint{}

\title{Curvature Instability of Membranes near Rigid Inclusions}



\author{S.\ Alex Rautu}
\email{stefanar@ncbs.res.in}
\affiliation{Simons Centre for the Study of Living Machines, National Centre for Biological Sciences (TIFR), Bellary Road, Bangalore 560065, India}
%\thanks{}

%\collaboration{}%\noaffiliation

\date{\today}

\begin{abstract}
In multicomponent membranes, internal scalar fields may couple to membrane curvature, thus renormalizing the membrane elastic constants and destabilizing the flat membranes. Here, a general elasticity theory of membranes is considered that employs a quartic curvature expansion. The shape of the membrane and its deformation energy near a long rod-like inclusion are studied analytically. In the limit where one can neglect the end-effects, the nonlinear response of the membrane to such inclusions is found in exact form. Notably, new shape solutions are found when the membrane is curvature unstable, manifested by a negative rigidity. Near the instability point (i.e.\ at vanishing rigidity), the membrane is stabilized by the quartic term, giving rise to a new length scale and new scale exponents for the shape and its energy profile. The contact angle induced by an applied force at the inclusion provides a method to experimentally determine the quartic curvature modulus.

\end{abstract}


%\keywords{}

\maketitle


Each living cell, including their organelles, is bounded by a sac-like membrane that plays an active and crucial role in almost every cellular process \cite{Alberts2008}. In its most basic form, a biomembrane consists of a bilayer lipid structure that acts as a platform for a myriad of other biological macromolecules \cite{Singer1972, Engelman2005}. Especially, a multitude of proteins can be incorporated into (or absorbed onto) membranes, resulting in a number of biological functions~\cite{Alberts2008}. Despite their complexity, biomembranes show a clear separation of scales due to their small thickness (about 5~nm) in comparison with their lateral extent (50~nm to 100~$\mu$m). This suggest that the large-scale properties of biomembranes may be adequately described by two-dimensional elastic sheets, controlled by bending rigidities and surface tension \cite{Helfrich1973,Canham1970,Evans1974}. Such a large-scale theory can be constructed through an effective free-energy defined on a surface $\mathcal{S}$~\cite{Deserno2015}. In particular, the membrane deformations near rigid inclusions (such as ion channels, pore-forming toxins,
and protein coats) can be prescribed by a few local fields that live on this surface domain $\mathcal{S}$~\cite{Phillips2009}, e.g.\ the mid-plane of the bilayer, the membrane thickness, the lipid tilt, and the relative concentration across the lipid bilayer \cite{Dan1993, Fournier1999, Weikl1998, Kim1998, Goulian1993, Wiggins2005, Rautu2015}. 

In the simplest case, the membrane elasticity can be described solely in terms of geometrical quantities, where the associated free-energy $\mathcal{H}$ is given by an expansion to second order in the surface invariants of $\mathcal{S}$, such as the area $\int\!\mathrm{d}S$, the mean curvature~$H$, and the~Gaussian curvature $K$, which yields the {\it Helfrich Hamiltonian} \cite{Deserno2015}:
\begin{equation}
  \label{eqn:free-energy}
  \mathcal{H}\left[\mathcal{S}\right]=\int_\mathcal{S}\mathrm{d}S\left[\sigma + \frac{\kappa}{2}\left(2 H-C_0\right)^2 + \bar{\kappa}\hspace{1pt}K\right]\hspace{-3pt},
\end{equation} where $\sigma$ is the surface tension, $C_0$ is the spontaneous curvature, $\kappa$ is the bending rigidity, and $\bar{\kappa}$ is the Gaussian curvature modulus. At equilibrium, the membrane shape $\mathcal{S}$ chooses the form that minimizes its associated free-energy in Eq.~(\ref{eqn:free-energy}) and subject to other constraints, if any (e.g.\ a fixed difference in area of the two membrane leaflets~\cite{Miao1994}, or a constant enclosed volume in the case of vesicles~\cite{Seifert1997}). This leads to an Euler-Lagrange equation (known as {\it the shape equation}), which is difficult to solve in general~\cite{Zhong-can1987}, and only a few exact solutions are known~\footnote{This includes Clifford tori~\cite{Zhong-can1990}, Dupin cyclides~\cite{Zhong-can1993}, noncircular cylinders~\cite{Shao-guang1996, Zhong-Can1999}, ellipsoids~\cite{Zheng1993, Liu1999, Naito1993}, Delaunay surfaces \cite{Naito1993, Jian-Guo1993, Mladenov2002, Naito1995}, and a few axisymmetrical vesicles~\cite{Naito1995}}. Besides, numerical methods have revealed more intricate surfaces, such as the biconcave shape of red blood cells \cite{Deuling1976, LimHW2002, Kralj-Iglic1993} and asymmetrical vesicles \cite{Kralj-Iglic1993, Heinrich1993, Jie1998}.

\begin{figure}[b]\includegraphics[width=0.6\columnwidth]{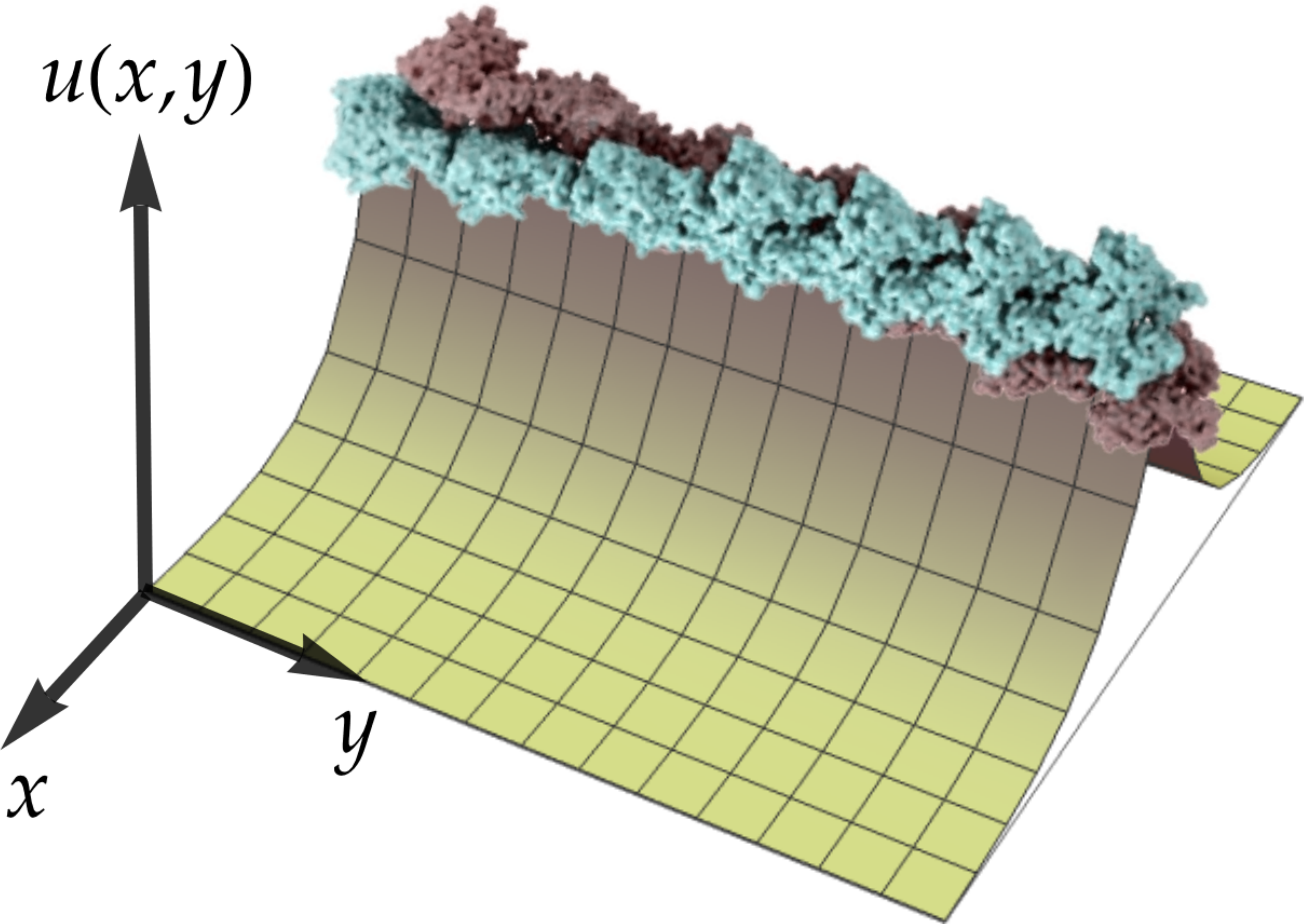}
\caption{\label{fig:1} Diagram of a biomembrane near a rigid inclusion (e.g.\ an actin bundle adsorbed on its surface) that is infinitely long in one direction, denoted by the $y$-axis. This induces a shape deformation along the $x$-axis, and its geometry is described within the Monge gauge, where its surface is given by a height function $u(x,y)$ above a flat reference plane.
}\end{figure}

In this Letter, a theoretical method is developed that allows us to determine the exact nonlinear response of a membrane to the inclusion of a long rigid object, such as an adsorbed actin bundle on the membrane, as shown in Fig.~\ref{fig:1}. Firstly, we show how this analytical approach is used to obtain an exact solution to the shape equation of a Monge parametrized membrane, based on the Helfrich Hamiltonian in Eq.~(\ref{eqn:free-energy}). This nonlinear solution allows one to compute properties of biological relevance, which are also experimentally measurable. This includes the membrane shape, and its deformation energy, beyond the linearized regime previously studied~\cite{Fournier1999, Weikl1998, Wiggins2005, Kim2000, Kim1998, Goulian1993, Haselwandter2012,Deserno2004, Muller2007, Hu2013, Rautu2015}. In the linearized case, the membrane is approximated as a small deviation from flatness, and its shape equation can be solved exactly \cite{Rautu2015}. However, the latter is inadequate to describe the deformations of highly curved membranes, which are ubiquitous in living cells \cite{Alberts2008}. Secondly, we show that other internal degrees of freedom, emerging in the case of multi-component membranes, can also be included in this framework. These can couple to the mean curvature, and may destabilize flat membranes \cite{Leibler1987, Rautu2015}. To further investigate the morphology of the membrane near this instability, a more general elasticity theory is considered, that employs a quartic curvature expansion \cite{Fournier1997b, Goetz1996, Capovilla2003}. In this case, the shape equation of a symmetric membrane can also be solved analytically, and notably it provides us with new exact solutions to a curvature unstable membrane, manifested by a negative renormalized rigidity. Lastly, an experimental method is proposed to measure the quartic curvature modulus. 

As depicted in Fig.\,\ref{fig:1}, the rigid inclusion is much longer in the $y$-axis than the $x$-axis, resulting in a membrane with translational invariance in the limit that the end-effects can be ignored. We seek its ground-state solution when the membrane is asymptotically flat. This gives the shape in terms of a height function $u(x)$ and the distance $x$ away from the inclusion \footnote{Similar solutions have been found by means of an arc-length parametrization of the tangent angle $\psi(s)$ \cite{Deserno2004, Muller2007, Hu2013}, which allows for overhangs. By integrating over $\sin\psi$ and $\cos\psi$, we attain the vertical and horizontal profiles of the membrane, respectively. However, one needs to numerically resolve for these to find the shape in terms of more relevant variables, e.g.\ the distance from the inclusion.}.
Although the system is effectively quasi-one-dimensional, it can be shown that its associated solution to the membrane shape yields the correct (and asymptotically exact) far-field behavior of a membrane deformed in response to a circular rigid inclusion~\footnote{For an axisymmetrical rigid inclusion~\cite{Rautu2015}, Eq.~(\ref{eqn:free-energy}) can~be written as $2\pi\!\int\!\mathrm{d}r\left[r f(r)\right]$, with $f$ defined in Eq.~(\ref{eqn:f}). The prefactor $r$ yields an extra term in the Euler-Lagrange equation, i.e.\ $g(r)\equiv r^{-1}[-\partial f/\partial u' + \frac{\mathrm{d}}{\mathrm{d}r}\!\left(\partial f/\partial u''\right)]$, where the remaining terms are identical to those found in the quasi-one-dimensional case. As $g(r)$ becomes negligibly small and asymptotically vanishes far from the inclusion, the leading terms yield a shape equivalent and asymptotically exact to Eq.~(\ref{eqn:ux}). Such axisymmetrical solutions have been studied in \cite{Biscari2007} using asymptotic methods.}, such as a transmembrane protein~\cite{Fournier1999}. 

Using Eq.~(\ref{eqn:free-energy}), the effective free-energy {\it per unit length} of such a quasi-one-dimensional membrane is given by  $\mathcal{F}\left[u(x)\right] = \int\mathrm{d}x\,f(x)$, with the free-energy density being
\begin{equation}
 \label{eqn:f}
 f(x) = \hat{\sigma}\sqrt{1\!+\!u'(x)^2} +\frac{C_0\,u''(x)}{1\!+\!u'(x)^2}+\frac{\frac{1}{2}\,\kappa\,u''(x)^2}{ \left[1\!+\!u'(x)^2\right]^{5/2}},
\end{equation} where the isotropic tension $\hat{\sigma}=\sigma+\kappa\,C^2_0/2$. Hereinafter, the dash and double-dash symbols denote the first and the second derivative with respect to the argument of the function, respectively. Eq.~(\ref{eqn:f}) is derived using that the area element is $\mathrm{d}S = \mathrm{d}x\,\mathrm{d}y\,\sqrt{1\!+\!u'(x)^2}$, the mean curvature is $H = -\frac{1}{2}\frac{\partial}{\partial x}\!\big[u'(x)/\sqrt{1\!+\!u'(x)^2}\,\big]$, and the Gaussian curvature $K=0$. By standard variational methods, the Euler-Lagrange equation of $\mathcal{F}[u(x)]$ is found to be~\cite{Note4}: $\frac{\mathrm{d}^4 u}{\mathrm{d}\hspace{0.5pt}x^4}\!=\!\frac{\left[1+u'\!(x)^2\right] u''\!(x)}{\lambda ^2}\!+\!\frac{5\left[1-6\, u'\!(x)^2\right] u''\!(x)^3}{2 \left[1+u'\hspace{-1pt}(x)^2\right]^2}\!+\!\frac{10\,u'\!(x)\,u''\!(x)}{1+u'\hspace{-1pt}(x)^2}\frac{\mathrm{d}^3 u}{\mathrm{d}\hspace{0.5pt}x^3},$ where $\lambda = \sqrt{\kappa/\hat{\sigma}\,}$. By defining $v(x) = u'(x)$ in the above equation, and then by assuming that $v'(x)$ is only a function of $v(x)$, i.e.\ $Q(v)=v'(x)^2$, this ansatz yields \footnote{Supplemental Material provides additional calculations, and includes Refs. \cite{Deserno2015, Rautu2015, Abramowitz1965}.}:
\begin{equation}
 \label{eqn:F-eqn}
 {Q}''(v) = \frac{2\left(1+v^2\right)}{\lambda ^2}+\frac{5 \left(1-6\hspace{1pt}v^2\right)\!
   {Q}(v)}{\left(1+v^2\right)^2}+\frac{10\hspace{1pt}v\hspace{1pt}{Q}'(v)}{1+v^2}.
\end{equation} This differential equation can be solved by a method of variation of parameters~\cite{Note4}, and the general solution is given by $Q(v) = 2\,\lambda^{-2}\left(1+v^2\right)^3 + \left(\mathcal{C}_1+v\,\mathcal{C}_2\right)\left(1+v^2\right)^{5/2}\!$, where $\mathcal{C}_1$ and $\mathcal{C}_2$ are constants of integration, which are fixed by the boundary conditions that the membrane becomes flat only at distances far from the inclusion. This asymptotic flatness can be imposed by requiring that both $Q(v)$ and its derivative $Q'(v)$ vanish in the limit of $v\rightarrow0$, which yields that $\mathcal{C}_1=-2\hspace{1pt}\lambda^{-2}$ and $\mathcal{C}_2=0$ \footnote{For the inner membrane between two inclusions, this does not hold, and one has to integrate first, to attain $u(v)$, and then apply the corresponding boundary conditions.}. 

\begin{figure}[t]\includegraphics[width=\columnwidth]{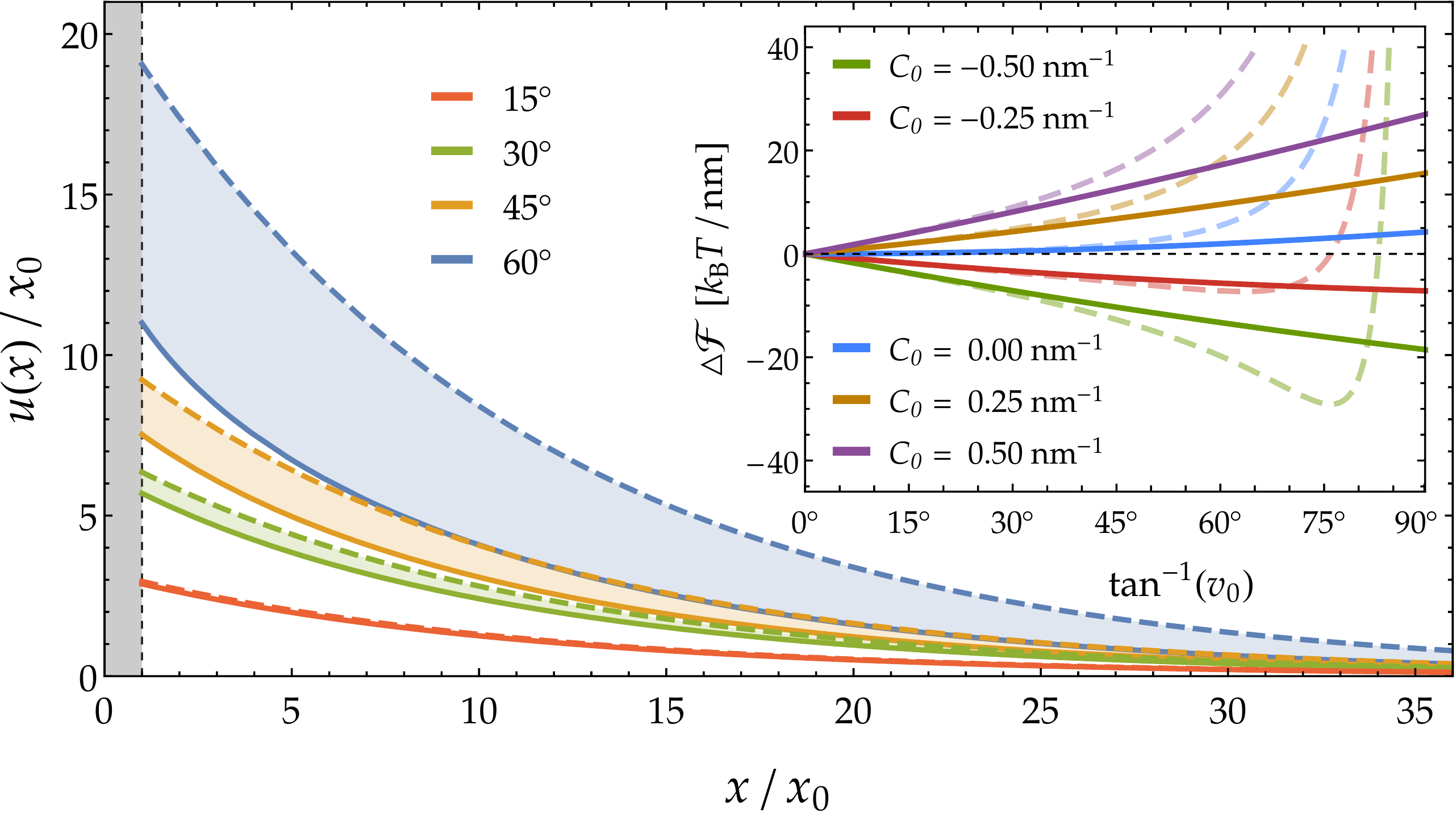}\caption{\label{fig:2} Membrane profiles near an inclusion for a few vaules of the contact angles at vanishing spontaneous curvature ($C_0=0$). Here, we take $x_0=1$\,nm, $\kappa=20\,k_B T$, and $\sigma=0.5$\,mN/m. The solid curves are the exact solutions from Eq.~\ref{eqn:ux}, while the dashed curves are their corresponding linearized versions. The inset plot shows the deformation free-energy per unit length $\Delta\mathcal{F}$ against the angles $\tan^{-1}\!\left(v_0\right)$ for a few values of $C_0$, where the solid line is the result in Eq.~\ref{eqn:F-0}, whilst the dashed curve is the associated linearized energy.}\end{figure}

By chain rule, we have that $v(x) = v'(x)\,u'(v)$, then the membrane height as a function of $v$ is found to be
\begin{equation}
 \label{eqn:uv}
 u(v) = \pm\int\limits^{\,v}_0\!\mathrm{d}\nu\,\frac{\nu}{\sqrt{Q(\nu)\,}} = \pm\lambda\,\sqrt{2-\frac{2}{\displaystyle\sqrt{1+v^2\,}}},
\end{equation} with the boundary condition $u\left(v\rightarrow0\right)=0$. Also, at the interface between the rigid inclusion and the membrane, we set the contact angle to be given by $\vartheta\equiv\tan^{-1}\!\left(v_0\right)$, which fixes the height at the inclusion to be $u_0 = u(v_0)$. The choice of what determines $u_0$ or $\vartheta$ is discussed later in this Letter. Now, by inverting Eq.~(\ref{eqn:uv}), $v$ is given by
\begin{equation}
 \label{eqn:vx}
 v(u)=\pm\,u\hspace{0.5pt}\sqrt{4\hspace{1pt}\lambda^2-u^2}\hspace{1pt}/\left(2\hspace{1pt}\lambda^2-u^2\right) = u'(x),
\end{equation} where $\left|u\right|<\lambda\sqrt{2}$ is required, setting an upper bound for the height function. Since the region spanned by the inclusion is chosen to be the interval $\left[-x_0,\,x_0\right]$, with $x_0\geq0$, then Eq.~(\ref{eqn:vx}) can be integrated over $u$ to find the position of the outer membrane $x(u) = \pm\big[x_0+\mathcal{X}(u_0)-\mathcal{X}(u)\big]$, with $\mathcal{X}(u) = \sqrt{4\hspace{1pt}\lambda^2-u^2}-\lambda\,\text{arccosh}\hspace{-0.5pt}\left(2\hspace{1pt}\lambda/\hspace{-0.5pt}\left|u\right|\right)$. The minus sign represents the negative regime with $x\in[-x_0,-\infty)$, whilst the plus sign corresponds to $x\in[x_0,\infty)$. Thus, the solution for the membrane height can be written as
\begin{equation}
 \label{eqn:ux}
 \left|u\right| = 2\hspace{1pt}\lambda\,\text{sech}\!\left[\!\sqrt{4-u^2/\lambda^2}  - \left[x_0 - \left|x\right|+\mathcal{X}(u_0)\right]\!\hspace{0.5pt}/\hspace{0.5pt}\lambda\right]\!\!,
\end{equation} with sech as the hyperbolic secant \cite{Abramowitz1965}. Typical membrane profiles are plotted in Fig.~\ref{fig:2}. The linearized solution can be retrieved by expanding Eq.~(\ref{eqn:ux}) to lowest order in $u$, yielding $u(x)\simeq \pm u_0\,e^{-\left(|x|-x_0\right)/\lambda}$ and $u_0\simeq v_0\hspace{0.5pt}\lambda$.  This allows us to estimate an upper bound for the contact angle at which the linearized solution is a good approximation, which is found to be about $10^\circ\hspace{-0.75pt}$ and $20^\circ\hspace{-0.75pt}$ at 1\% and 5\% maximal relative error, respectively~\footnote{This relative error is independent of $\lambda$, and can be exactly derived from Eq.~(\ref{eqn:uv}), i.e.\ $\Delta\mathcal{E} = v_0/\sqrt{2-2/\sqrt{1+v_0^2\,}}-1$.}.

Once the membrane shape is determined, the deformation energy can be computed analytically. The energetic cost per unit length to deform a membrane from flatness is given by $\Delta\mathcal{F} \equiv \mathcal{F}\left[u\right]-\mathcal{F}\left[0\right] = 2\int^{\infty}_{x_0}\mathrm{d}x\left[f(x)-\hat{\sigma}\right]$.~By a change of variables and using Eq.~(\ref{eqn:vx}), this yields \cite{Note4}:
\begin{equation}
\label{eqn:F-0}
 \Delta\mathcal{F} = 4\sqrt{\kappa\hspace{1pt}\hat{\sigma}\hspace{0.5pt}}\left[2-\sqrt{4-\frac{u_0^2}{\lambda^2}}+\lambda\hspace{1pt}C_0\hspace{1pt}\text{csc}^{-1}\!\left(2\hspace{1pt}\lambda/u_0\right)\right]\!\!,
\end{equation} where csc$^{-1}$ is the inverse cosecant. In the small angle approximation (linearized case), the deformation energy reduces to $\Delta\mathcal{F}\simeq\sqrt{\kappa\hspace{1pt}\hat{\sigma}\hspace{0.5pt}}\left(v_0^2+2\hspace{1pt}v_0\hspace{0.5pt}\lambda\,C_0\right)\hspace{-0.5pt}$, as $u_0\simeq v_0\hspace{0.5pt}\lambda$. The dependences of $\Delta\mathcal{F}$ on $v_0$ and $C_0$ are shown in Fig.~\ref{fig:2}.

\begin{figure}[t]\includegraphics[width=\columnwidth]{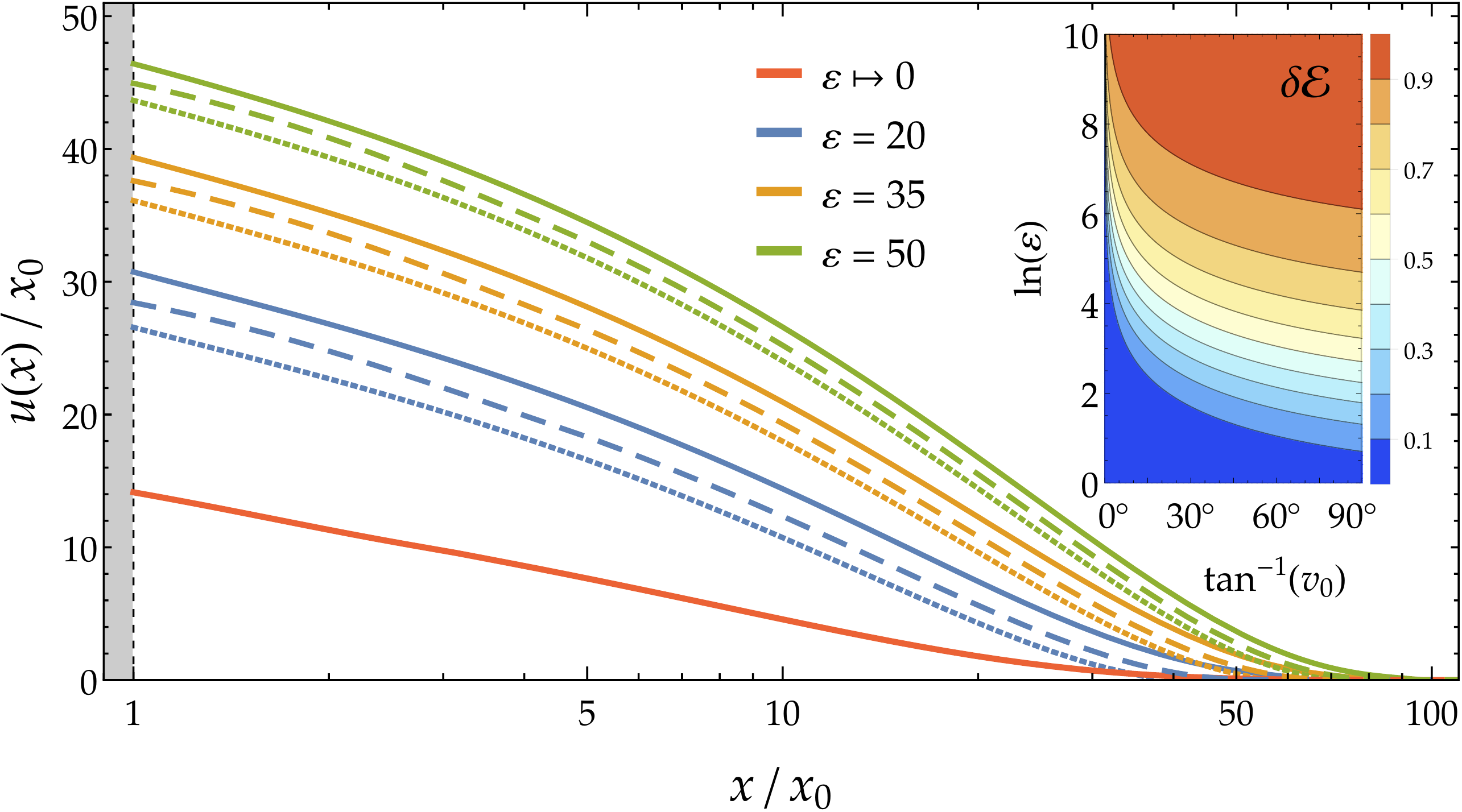}\caption{\label{fig:3} Semi-log plot of the membrane deformation profiles for a few values of $\varepsilon$ (see Eq.~\ref{eqn:uv-4}). The solid and dotted lines correspond to the regimes of $\tilde{\kappa}>0$ and $\tilde{\kappa}<0$, respectively. The dashed curves are  the asymptotic solutions as $\varepsilon\!\rightarrow\!\infty$ (see Eq.~\ref{eqn:uv-asymptotic}). Herein, the contact angle is $80^{\circ}$, while $x_0=1$\,nm, $\sigma=0.5$\,mN/m, and $\left|\tilde{\kappa}\right|\hspace{-0.5pt}=\hspace{-0.5pt}20\,k_B T$. The inset shows the relative error $\delta\mathcal{E}\!=\left(u[\varepsilon]-u[\varepsilon=0]\right)\!/u[\varepsilon]$, with $\tilde{\kappa}>0$ and $x=x_0$, as a function of $\ln(\varepsilon)$ and the contact angle.}\end{figure}

So far, only the energetics incurred by membrane shape deformations have been considered. However, other fields that live on the surface $\mathcal{S}$ can be taken into account  \cite{Leibler1987}. In the case of a mixed two-component lipid membrane, the local relative concentration between the leaflets of the membrane, say $\varphi$, incurs an energetic contribution of the form $\mathcal{F}_\varphi[\mathcal{S}] = \int_\mathcal{S}\mathrm{d}{S}\left[\frac{1}{2}\hspace{1pt}a\hspace{1pt}\varphi^2 + 2\hspace{1pt}c\hspace{1pt}\varphi H\right]\!$ to lowest order in $\varphi$, where $a$ and $c$ are phenomenological parameters~\cite{Rautu2015}. By minimizing over $\mathcal{F}+\mathcal{F}_\varphi$, the ground state solution to $\varphi$ for a quasi-one-dimensional membrane (as shown in Fig.~\ref{fig:1}) is  $\varphi=-2\hspace{1pt}c\hspace{1pt}H/a$ \cite{Note4}. This effectively renormalizes $\kappa$ to a new bending modulus $\tilde{\kappa} = \kappa\hspace{1pt}[1-c^2/\left(a\kappa\right)]$, and thus the corresponding membrane height is found to be in a form identical to Eq.~(\ref{eqn:ux}) where $\lambda\mapsto\tilde{\lambda}=\!\sqrt{\tilde{\kappa}/\hat{\sigma}\hspace{1pt}}$. The regime $c>\!\sqrt{a\kappa}$ corresponds a membrane curvature instability, know as the Leibler's unstable regime \cite{Leibler1987}.

The free-energy functional in Eq.~(\ref{eqn:free-energy}) is found by truncating at second-order in curvature, and assuming that higher order terms are much smaller; however, {\it a priori} there is no clear basis for this assumption \cite{Deserno2015}. In fact, higher-order terms become more important as the Leibler instability is approached, leading to a much softer bending modulus. To better understand its physical implications, we consider a symmetric membrane and quasi-one-dimensional. The succeeding terms that appear at fourth-order in the inverse-length are $K^2\!$, $K\hspace{0.25pt}H^2\!$, $H^4\!$, and $\left(\nabla_\alpha H\right)\!\left(\nabla^\alpha H\right)$~\cite{Capovilla2003}. As $K=0$, the first two terms vanish identically. The last term is neglected henceforth, as it contains derivatives of mean curvature, introducing sub-dominant terms, and thus we only consider the $H^4$ term. 

Using the techniques developed previously, an exact solution to the shape can be found again for this quartic curvature theory of membranes, where the free-energy is $\mathcal{H}_4[\mathcal{S}] = \int_\mathcal{S}\mathrm{d}S\left[\sigma+2\hspace{1pt}\kappa\hspace{1pt}H^2+4\hspace{1pt}\kappa_4\hspace{0.5pt}H^4\right]\hspace{-1pt}$, with $\kappa_4\geq0$ as the quartic curvature modulus. As in Eq.~(\ref{eqn:f}), the free-energy per unit length is given by $\mathcal{F}_4\left[u(x)\right] = \int\mathrm{d}x\,f_4(x)$, where
\begin{equation}
 \label{eqn:f-4}
 f_4 = \sigma\sqrt{1\!+\!u'(x)^2} +\frac{\frac{1}{2}\,\kappa\,u''(x)^2}{ \left[1\!+\!u'(x)^2\right]^{\frac{5}{2}}}+\frac{\frac{1}{4}\,\kappa_4\,u''(x)^4}{ \left[1\!+\!u'(x)^2\right]^{\frac{11}{2}}}
\end{equation} represents the projected free-energy density. The shape equation of $\mathcal{F}_4+\mathcal{F}_\varphi$ can be found \cite{Note4} and rewritten in terms of a new function $\mathcal{Q}(v)=v'(x)^2$ as before, where~$v$ is the membrane gradient. By the condition~of asymptotic flatness, both $\mathcal{Q}(v)$ and $\mathcal{Q}'(v)$ are required to vanish at $v=0$, which yields the following two solutions \cite{Note4}:
\begin{equation}
 \label{eqn:Q-4}
 \mathcal{Q}_\pm(v) = \frac{4\hspace{1pt}\sigma\left(1+v^2\right)^3}{\pm\hspace{1pt}\tilde{\kappa}\hspace{1pt}\hspace{1pt}\varepsilon^2}\!\left[\sqrt{1+\varepsilon^2-\frac{\varepsilon^2}{\sqrt{1+v^2}}\,}\mp1\right]\!\!, 
\end{equation} where $\varepsilon =\xi\sqrt{12}/\tilde{\lambda}$, with $\xi=\sqrt{\kappa_4/|\tilde{\kappa}|}$ and $\tilde{\lambda}=\sqrt{|\tilde{\kappa}|/\sigma}$. As~$\mathcal{Q}$ must be greater than zero, only $\mathcal{Q}_+$ solution is allowed if $\tilde{\kappa}>0$. However, if the bending modulus switches sign and becomes negative, as in the case of the Leibler's instability, then only $\mathcal{Q}_-$ must be chosen for stability.

By the flatness condition $u(v\rightarrow0)=0$, c.f.~Eq.~(\ref{eqn:uv}), the profile $u$ can be exactly computed in terms of $v$~\cite{Note4}:
\begin{equation}
 \label{eqn:uv-4}
 \left|u(v)\right|=U_\pm+\frac{\tilde{\lambda}^2\sqrt{\mathcal{Q}_\pm(v)\,}}{\left(1+v^2\right)^{3/2}}\left[\frac{\varepsilon^2\tilde{\lambda}^2 \mathcal{Q}_\pm(v)}{12 \left(1+v^2\right)^3}\pm1\right]\!\!,
\end{equation} with $U_+\!=0$ and $U_-\!=2\tilde{\lambda}\sqrt{2}/(3\varepsilon)$. For $\tilde{\kappa}>0$, Eq.~(\ref{eqn:uv}) is retrieved when $\varepsilon\!\rightarrow0$ ($\tilde{\lambda}\gg\xi$), whereas for $\tilde{\kappa}<0$ the height $u$ vanishes at $\varepsilon=0$ and its linear order term in $\varepsilon$ gives $\left|u(v)\right|\!=\!\xi\left(1-1/\sqrt{1+v^2}\right)\!\sqrt{3/2}$. On the other hand, in the asymptotic limit $\varepsilon\!\rightarrow\!\infty$ ($\tilde{\lambda}\ll\xi$), i.e.\ near the instability, both solutions yield the same profile:
\begin{equation}
 \label{eqn:uv-asymptotic}
 \left|u(v)\right|\simeq 2^{-3/4}\,\Lambda\left(1-1/\sqrt{1+v^2}\,\right)^{3/4}
\end{equation} with $\Lambda=\frac{4}{3}\sqrt[4]{6\hspace{1pt}\kappa_4/\sigma}$ \cite{Note4}. Then, by integrating over $u$, the membrane position from the rigid inclusion (namely, $|x|\geq x_0$ and $|u|\leq|u_0|$) is found to be~\cite{Note4}: $|x(u)| \simeq x_0 + \mathcal{X}_\infty(u_0) - \mathcal{X}_\infty(u)$, where $\mathcal{X}_\infty(u)\!=\!\frac{\Lambda}{2}\,\text{sn}^{\text{--}1}\big(\!\sqrt[3]{|u|/\Lambda}\,\big|\!-\!1\big)+\Lambda\sqrt{(u/\Lambda)^{2/3}-(u/\Lambda)^{2}}$, with $sn^{\text{--}1}\big(\vartheta\hspace{0.5pt}| m\big)$ as the inverse of the $sn$ Jacobi elliptic function~\footnote{This is defined by $sn^{\textit{-}1}(z\,| m) = \int^{z}_0\!\frac{\mathrm{d}t}{\sqrt{1-t^2}\sqrt{1-m t^2}}$, where the real parameters $z\in\!\left(\textit{--}1,1\right)$ and $m\in\!\left(\textit{--}\hspace{1pt}\infty,1\right)$.}. Therefore, this gives the height in terms of $|x|$ as shown in Fig.~\ref{fig:3}~\footnote{For $\xi\gg\tilde{\lambda}$, the small gradient $\frac{\mathrm{d}u}{\mathrm{d}x}\sim u^{2/3}$ that yields the algebraic solution $x(u)\sim u^{1/3}$. On the other hand, we have $\frac{\mathrm{d}u}{\mathrm{d}x}\sim u$ for $\tilde{\lambda}\gg\xi$, which leads to $x(u)\sim\log \left|u\right|$.}. 

In its exact form, Eq.~(\ref{eqn:uv-4}) can be expressed as a cubic equation in $\mathcal{V}\equiv1-1/\sqrt{1+v^2}$, which has only one real root. Consequently, Eq.~(\ref{eqn:uv-4}) can be analytically inverted for any $\varepsilon$, but now the inverse gradient, $1/v(u)$, has a cumbersome expression, and its integral over $u$ cannot be written in a closed form \cite{Note4}. Nonetheless, this can be numerically integrated (see  Fig.~\ref{fig:3}). For $\tilde{\kappa}>0$, the maximal relative error $\delta\mathcal{E}$ in neglecting the quartic term can be exactly computed from Eq.~(\ref{eqn:uv-4}), and its result is shown in the inset plot of Fig.~\ref{fig:3}. If $\xi=\tilde{\lambda}$, then we need a contact angle of about $14^\circ$ for $\delta\mathcal{E}=10\%$.

\begin{figure}[t]\includegraphics[width=\columnwidth]{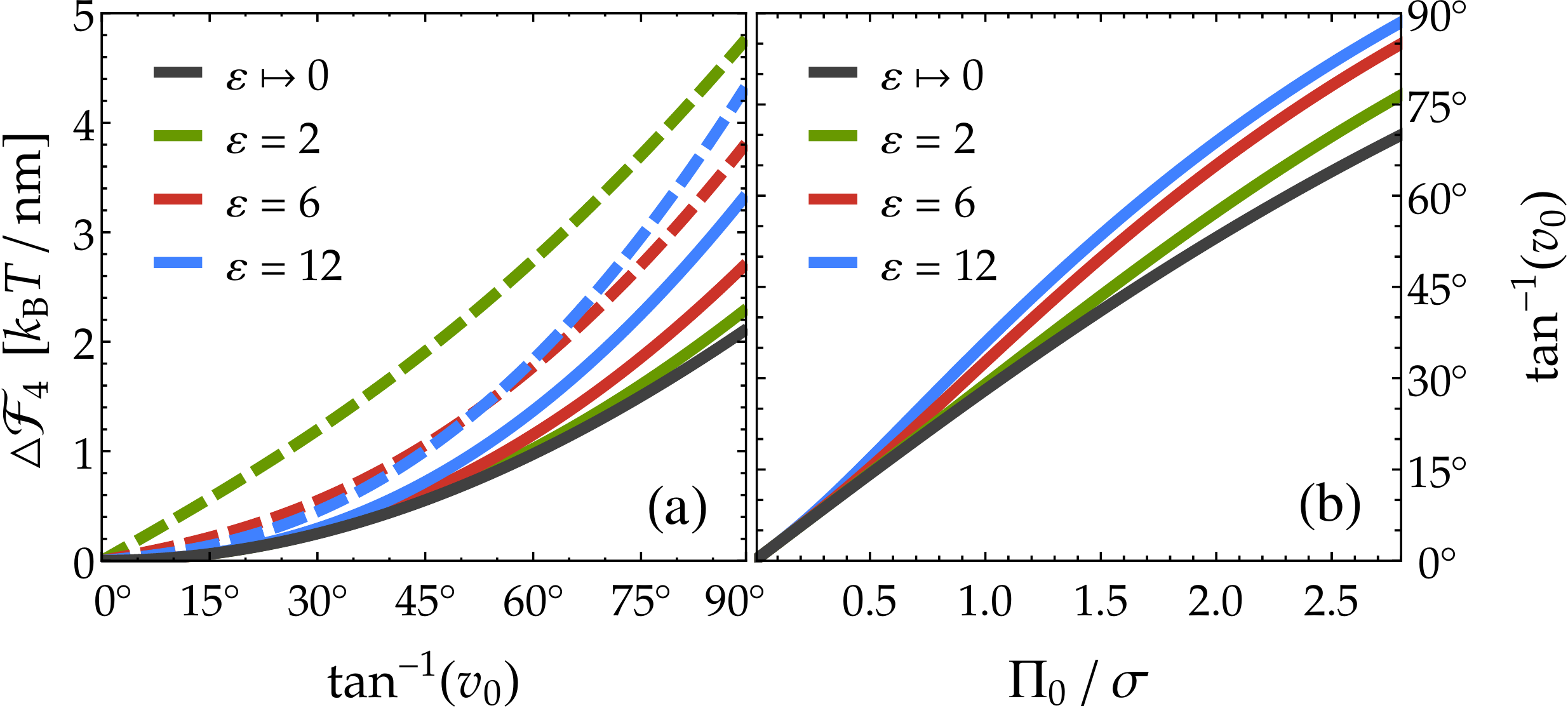}\caption{\label{fig:4} (a) The deformation free-energy per unit length, $\Delta\mathcal{F}_4$, in the quartic curvature theory of membranes (see main text). By setting $\sigma=0.5$\,mN/m, $\Delta\mathcal{F}_4$ versus the contact angle is shown for a few values of $\varepsilon$ when $\tilde{\kappa}=20\,k_B T$ (solid lines) and $\tilde{\kappa}=-25\,k_B T$ (dashed lines). (b) The induced contact angle $\tan^{-1}\!\left(v_0\right)$ at the inclusion--membrane interface as a function of the pulling force per unit length $\Pi_0$ on the rigid inclusion, that is normalized by the surface tension $\sigma$.}\end{figure}

Using Eqs.~(\ref{eqn:Q-4}) and (\ref{eqn:uv-4}) for $\tilde{\kappa}>0$, the deformation free-energy from flatness $\Delta\mathcal{F}_4$ can be reduced to~\cite{Note4}:
\begin{equation}
 \Delta\mathcal{F}_4 = 2\hspace{1pt}\sigma\int^{v_0}_0\frac{u(v)}{1+v^2}\,\mathrm{d}v,
\end{equation} which can be numerically integrated, see Fig.~\ref{fig:4}\,(a). By expanding in $\varepsilon$ about zero, this gives the lowest order contribution due to the quartic term, namely $\Delta\mathcal{F}_4 = \Delta\mathcal{F}+\frac{\tilde{\kappa}\varepsilon^2}{18\tilde{\lambda}}\Big[8-\big(5-1/\sqrt{1+v_0^2\,}\,\big) \sqrt{2+2/\sqrt{1+v_0^2\,}}\,\Big] + \mathcal{O}[\varepsilon^4]$, where $\Delta\mathcal{F}$ is given by Eq.~(\ref{eqn:F-0}) with $\kappa\mapsto\tilde{\kappa}$ and $C_0=0$. This result can be used to determine the values of $\xi$ at which the $\varepsilon^2$-term becomes greater than $\Delta\mathcal{F}$. Thus, by equating these two terms, we find at the modest contact angles of $15^\circ\hspace{-0.75pt}$ and $30^\circ\hspace{-0.75pt}$ that $\xi\approx11\hspace{1pt}\tilde{\lambda}$ and $\xi\approx5\hspace{1pt}\tilde{\lambda}$, respectively, showing that the quartic terms in the mean curvature can dominate at larger contact angles ($\gtrsim\!45^\circ$). On the other hand, the deformation free-energy for $\tilde{\kappa}<0$ takes a different form than before~\cite{Note4}, which can be written as $\Delta\mathcal{F}_4 = \sigma\int^{v_0}_0\mathrm{d}v\,\frac{\varepsilon^2\tilde{\lambda}^4\mathcal{Q}_-(v)^{3/2}}{6\hspace{1pt}\left(1+v^2\right)^{11/2}}$, and is shown in Fig.~\ref{fig:4}\,(a). Hence, its expansion in~$\varepsilon$ about zero is given by $\Delta\mathcal{F}_4 = \frac{4\tilde{\lambda}}{3\hspace{1pt}\varepsilon\sqrt{2}}\tan^{\text{--}1}(v_0) + \frac{\varepsilon\tilde{\lambda}}{2\sqrt{2}}\Big[\!\tan^{\text{--}1}(v_0) -\frac{v_0}{\sqrt{1+v_0^2}}\Big]\! + \mathcal{O}[\varepsilon^3]$, which is linear in the contact angle for small $v_0$ or $\varepsilon$. 

However, near the instability point, $\tilde{\kappa}\rightarrow0$ (or $\varepsilon\rightarrow\infty$), we find the same expression for the both signs of $\tilde{\kappa}$~\cite{Note4}, i.e.\ $\Delta\mathcal{F}_4\simeq2\hspace{1pt}\sigma\int^{v_0}_0\!\mathrm{d}v\,\frac{u(v)}{1+v^2}+\mathcal{O}[\frac{1}{\varepsilon}] \simeq \frac{\sigma\Lambda\sqrt{2}}{5}\,v_0^{\frac{5}{2}}\! + \mathcal{O}[v_0^{\frac{9}{2}}]$, where the form of $u(v)$ from Eq.~(\ref{eqn:uv-asymptotic}) is implied, and the last equality is a small $v_0$ expansion. The latter shows a superquadratic dependence on $v_0$ of the incurred energy, in contrast to Eq.~(\ref{eqn:F-0}) at small angles where $\Delta\mathcal{F}\propto v^2_0$.

The boundary condition at the interface between the inclusion and the membrane has been specified by a fixed contact angle $\vartheta=\tan^{-1}\!\left(v_0\right)$~\footnote{This boundary condition could have also been specified by the value of the scalar field $\varphi$ at this interface, allowing the compositional field to modulate the contact angle.}. In the case of a protein assembly absorbed on the membrane, the nonzero value of  $\vartheta$ is a result of both the shape conformation of the rigid inclusion and its affinity to the membrane~\cite{Phillips2009}. Elongated protein structures, such as actin bundles and amyloid fibrils \cite{Pieters2016}, are examples of such inclusions. Due to cellular activity, they can also be pushed against the membrane, leading to an increase of $\vartheta$. In fact, this boundary condition can be recast as an applied force (per unit length) $\Pi_0$ that keeps the height $u_0=u(v_0)$ constant at $x=x_0$ along the $y$-axis. To find $u_0$, and hence $\vartheta$, we minimize the function $\mathcal{L}(u_0)\equiv\Delta\mathcal{F}_4-u_0\Pi_0$, where $\Pi_0$ acts as a Lagrange multiplier. Assuming that $\tilde{\kappa}>0$, the condition $\mathcal{L}'(u_0)=0$ leads to the following exact expression~\cite{Note4}:
\begin{equation}
 \label{eqn:load}
 \Pi_0 = \frac{8\hspace{0.5pt}\sigma\big{[}\varepsilon^2(1-\cos\vartheta) - 1 +\!\sqrt{\varepsilon^2(1-\cos\vartheta)+1}\,\big{]}}{3\hspace{0.5pt}\varepsilon^2\sin\vartheta},
\end{equation} which allows one to find the contact angle~$\vartheta$ induced by an external force $\Pi_0$ as shown in Fig.~\ref{fig:4}\,(b). Thus, this can be performed experimentally: an external load is applied on a flat membrane, and then its shape deformation and $\vartheta$ can be measured through an optical microscopy technique (e.g.\ confocal imaging). By keeping the membrane under a tension $\sigma$, and varying $\Pi_0$, an estimate of $\varepsilon$ (or $\kappa_4$ as $\kappa$ is typically known) is inferred by fitting to~Eq.~(\ref{eqn:load}). This gives us a simple method to find the quartic bending modulus $\kappa_4$, whose value is currently still unknown.

In summary, we computed exactly the energetics and shape of a membrane deformed by a rigid inclusion. We studied its nonlinear response in the Monge gauge by using the Helfrich Hamiltonian as well as a quartic curvature theory of membranes. The latter reveals interesting new physics; in particular, new shape solutions have been obtained above and below the curvature instability of membranes ($\kappa=0$). In the vicinity of this instability, both solutions yield the same morphology and deformation energy, which are fundamentally different to those found in stable membranes, with $\kappa>0$. Furthermore, the induced contact angle under an externally applied force can be calculated. This could plausibility be performed experimentally, allowing one to estimate the material parameters of membranes, such as the elusive value of the quartic curvature modulus.

\begin{acknowledgments}
The author acknowledges the stimulating discussions with Prof.\ G.\ Rowlands (University of Warwick, United Kingdom), Prof.\ M.\ Rao, Dr.\ R.\ Morris, and A.\ Singh (National Centre for Biological Sciences, India), and the funding from the Simons Foundation (United States).
\end{acknowledgments}

\end{document}